\documentclass[twocolumn,showpacs,prb,amsfonts,amsmath,amssymb,floatfix]{revtex4}
\usepackage{color}
\usepackage{hhline}
\usepackage{mathrsfs}
\usepackage{graphicx}
\usepackage{dcolumn}
\usepackage{bm}
\usepackage{multirow}
\usepackage{booktabs}
\usepackage{afterpage}
\usepackage{amsmath}

\begin{document}

\title{{\it Ab initio} GW calculation for organic compounds (TMTSF)$_2$PF$_6$}

\author{Kazuma Nakamura$^{1}$}
\author{Shiro Sakai$^{2}$}
\author{Ryotaro Arita$^{2}$}
\author{Kazuhiko Kuroki$^{3}$}
\affiliation{$^1$Quantum Physics Section, Kyushu Institute of Technology,
1-1 Sensui-cho, Tobata, Kitakyushu, Fukuoka, 804-8550, Japan}
\affiliation{$^2$Department of Applied Physics, The University of Tokyo, 
             Hongo, Bunkyo-ku, Tokyo 113-8656, Japan}
\affiliation{$^3$Department of Physics, Osaka University, 1-1 Machikaneyama, Toyonaka, Osaka 560-0043, Japan}

\date{10 April, 2013}
\begin{abstract}
We present an {\it ab initio} GW calculation to study dynamical effects on an organic compound (TMTSF)$_2$PF$_6$. Calculated polarized reflectivities reproduce experimental plasma edges at around 0.2 eV for $E\|b'$ and 1.0 eV for $E\|a$. The low-energy plasmons come out from the low-energy narrow bands energetically isolated from other higher-energy bands, and affect the low-energy electronic structure via the GW-type self-energy. Because of the quasi-one-dimensional band structure, a large plasmon-induced electron scattering is found in the low-energy occupied states along the X-M line.
\end{abstract}
\pacs{71.15.Mb, 71.45.Gm, 71.20.Rv}

\maketitle

\section{Introduction}\label{sec:intro} 
Physics and chemistry of organic conductors have recently attracted much attention owing to its low dimensionality, strong electron correlation, and material variety and flexibility. Despite the complicated structure of the molecules themselves, rather simple and energetically isolated band structure commonly seen in these materials provides ideal basis for studying fundamental physics of electron correlation. Among a number of organic conductors, (TMTSF)$_2$PF$_6$ has been of particular interest ever since its discovery as the first organic superconductor.~\cite{Bechgaard} In this material, the molecules are stacked along the $a$-axis, and this is the most conductive axis because the molecular orbitals are elongated in the stack direction. A small overlap between neighboring molecular orbitals in the interstack direction gives rise to a weak two-dimensionality. The importance of electron correlation can readily be seen from the presence of the spin density wave (SDW) phase, 
 which takes place below 12K at ambient pressure.~\cite{Ishiguro-Yamaji-Saito}  The SDW transition temperature decreases upon applying hydrostatic pressure, and superconductivity sits next to the SDW phase in the temperature-pressure phase diagram. The superconducting state also shows some interesting features suggesting unconventional pairing, which may also be a manifestation of electron correlation.~\cite{Kuroki}

Theoretically, the microscopic origin of the density waves and superconductivity has been intensively investigated on simplified Hubbard type models where on-site and/or short-ranged off-site repulsions are taken into account.~\cite{Kuroki} On the other hand, recent {\it ab initio} studies on organic materials show that the long-range part of the Coulomb interactions is appreciably present,~\cite{Nakamura-organic-1,Nakamura-organic-2} suggesting that dielectric properties can also be of interest. In fact, reflectance measurements of (TMTSF)$_2$PF$_6$ show presence of plasma edge,~\cite{Dressel,Jacobsen} indeed indicating the importance of the long range Coulomb interaction.

In general in solids, the energy scale of the plasmon excitation is of the order of 10 eV and therefore it is believed that such excitations are irrelevant to the low-energy physics of the order of 0.1-1 eV. However, in the organic materials, low-energy bands around the Fermi level tend to be isolated from other high-energy bands, resulting in a plasmon characterized by the bandwidth and occupancy of the isolated low-energy bands. Since their bandwidth is typically of the order of $\sim$1 eV, the plasma frequency can also be in this energy scale. Then, the plasmon excitation may produce new aspects in the low-energy electronic states through self-energy effects. 
Since recent progress in angle-resolved photo-emission spectroscopy~\cite{ARPES-1,ARPES-2,ARPES-3} has made it possible to measure quasiparticle band structure and to perform detailed self-energy analyses, corresponding first-principle calculations are highly desired. 

In the present study, we present an {\it ab initio} GW calculation to study dynamical effects on the electronic structure of the organic compound (TMTSF)$_2$PF$_6$. The GW calculation takes into account the effect of plasmon excitation.~\cite{Hedin,Hybertsen,Aryasetiawan,Onida,GW-1,GW-2,GW-3,GW-4,GW-5,GW-6,GW-7,GW-8,GW-9,GW-10,GW-11,GW-12,GW-13,GW-14,GW-15,GW-16,GW-17,GW-18,GW-19,GW-20,Nohara} 
The calculated reflectances well reproduce experimental results, identifying the experimentally-observed plasma edges to the plasmons within the low-energy bands. By calculating GW self energy and spectral function, we will show that this low-energy plasmon excitation affects the low-energy electronic structure. 
Since the isolated band character inducing the low-energy plasmon is ubiquitous in strongly-correlated electron systems such as organic compounds and transition-metal compounds, the present result will provide a general basis for analyzing various correlated materials.

\section{Method}\label{sec:method}
Here we describe our scheme. The non-interacting Green's function is given by  
\begin{eqnarray}
G_0({\bf r,r'},\omega) = \sum_{\alpha{\bf k}} 
\frac{\psi_{\alpha{\bf k}}({\bf r}) \psi^{*}_{\alpha{\bf k}}({\bf r'})}
{\omega-\epsilon_{\alpha{\bf k}}+i \delta {\rm sgn}(\epsilon_{\alpha{\bf k}}-\epsilon_f)}, 
\end{eqnarray}
where $\psi_{\alpha {\bf k}} ({\bf r})$ and $\epsilon_{\alpha {\bf k}}$ are the Kohn-Sham (KS) wavefunction and its eigenvalue of band $\alpha$ and wavevector ${\bf k}$, and $\epsilon_f$ is the Fermi level. $\delta$ is chosen to be a small but finite positive value to stabilize numerical calculations. A polarization function of a type $-i G_0 G_0$ is written in a matrix form in the plane wave basis as
\begin{eqnarray}
\chi_{{\bf GG'}}&({\bf q},\omega)&
\!=\!2\sum_{{\bf k}}\sum^{vir}_{\alpha}\sum^{occ}_{\beta}
M_{\alpha\beta}^{{\bf G}}({\bf k,q}) 
M_{\alpha\beta}^{{\bf G'}}({\bf k,q})^{*} \nonumber \\ 
\!\!\!&\times&\!\!\!\Biggl\{\!\!
\frac{1}{\omega\!-\!\epsilon_{\alpha{\bf k\!+\!q}}\!+\!\epsilon_{\beta{\bf k}}\!+\!i\delta}\!-\!
\frac{1}{\omega\!+\!\epsilon_{\alpha{\bf k\!+\!q}}\!-\!\epsilon_{\beta{\bf k}}\!-\!i\delta}\!\!\Biggr\} 
\label{eq:chi} 
\end{eqnarray}
with $M_{\alpha\beta}^{{\bf G}}({\bf k,q})=\langle \psi_{\alpha {\bf k+q}}|e^{i({\bf q+G}){\bf r}}|\psi_{\beta {\bf k}} \rangle$.

Optical properties in a metal are related to the symmetric dielectric function~\cite{Hybertsen} in the ${\bf q}\to0$ limit   
\begin{eqnarray}
\epsilon_{{\bf G\!G'}}(\omega)\!=\!\delta_{{\bf G\!G'}}\!-\!\frac{(\omega_{pl,\mu\mu})^2}{\omega(\omega\!+\!i\delta)}\delta_{{\bf G0}}\delta_{{\bf G'0}}\!-\!\!\lim_{{\bf q}\to0}\!\frac{4\pi}{N\Omega}\!\frac{\chi_{{\bf GG'}}^{inter}({\bf q},\omega)}{|{\bf q\!+\!G}||{\bf q\!+\!G'}|} \nonumber \\ 
\label{eq:eps}
\end{eqnarray}
with ${\bf q}$ approaching zero along the Cartesian $\mu$ direction. $N$ is the total number of sampling $k$ points and $\Omega$ is the unitcell volume. 
The second term is the Drude term due to the intraband transition around the Fermi level. The third term represents the interband contribution, where $\chi_{{\bf GG'}}^{inter}({\bf q},\omega)$ is a polarization matrix due to the interband transitions. Plasma frequency in the second term is given in a tensor form by~\cite{Draxl-1,Draxl-2}
\begin{eqnarray}
\omega_{pl,\mu\nu}=\sqrt{\frac{8\pi}{\Omega N} \sum_{\alpha{\bf k}} p_{\alpha {\bf k},\mu} p_{\alpha {\bf k},\nu} \delta(\epsilon_{\alpha {\bf k}}-\epsilon_f}) \label{wpl} 
\end{eqnarray}
with $p_{\alpha{\bf k},\mu}$ being a matrix element of a momentum as 
\begin{eqnarray}
p_{\alpha {\bf k},\mu} = - i \langle \psi_{\alpha {\bf k}}| \frac{\partial}{\partial x_{\mu}}+[V_{NL}, x_{\mu}] |\psi_{\alpha {\bf k}} \rangle, 
\label{p_ij} 
\end{eqnarray}
where $V_{NL}$ is the non-local part of the pseudopotential.

The GW self-energy is given by
\begin{eqnarray}
\Sigma({\bf r,r'},\omega)\!=\!i\!\int \frac{d\omega'}{2\pi} G_0({\bf r,r'},\omega+\omega') W({\bf r,r'},\omega'). 
\label{GW_SIGMA}
\end{eqnarray}
The screened Coulomb interaction $W(\omega)=v^{\frac{1}{2}}\epsilon^{-1}(\omega)v^{\frac{1}{2}}$ is decomposed into the bare Coulomb interaction $v$ and the frequency-dependent part $W_C(\omega)=W(\omega)-v$. The frequency integral of $iG_0 v$ gives the bare exchange term $\Sigma_X$ and that of $iG_0 W_C$ gives the correlation term $\Sigma_C(\omega)$ including the retardation effect. The calculation of $\Sigma_X$ is straightforward while that of $\Sigma_C(\omega)$ is somewhat technical. In the present calculation, we fit the following function to {\it ab initio} $W_C(\omega)$,~\cite{Nohara}  
\begin{eqnarray}
\tilde{W}_C({\bf r,r'},\omega) 
= \sum_{j} \Biggl( \frac{1}{\omega-z_j} + \frac{1}{\omega+z_j} \Biggr) a_j( {\bf r,r'}), 
\end{eqnarray}
where $z_j$ and $a_j({\bf r,r'})$ are the pole and amplitude of the model interactions, respectively. Since the frequency-dependent part is decoupled from the amplitude in $\tilde{W}_C$, the frequency integral in $iG_0\tilde{W}_C$ can be analytically performed. The resulting matrix elements of $\Sigma_C(\omega)$ is 
\begin{eqnarray}
\langle \psi_{\alpha {\bf k}}| \Sigma_C(\omega) |\psi_{\alpha {\bf k}} \rangle\!\!=\!\!\sum_{jn{\bf q}} 
\frac{\langle \psi_{\alpha {\bf k}} \psi_{n {\bf k-q}}| a_j |\psi_{n {\bf k-q}} \psi_{\alpha {\bf k}} \rangle} 
{\omega\!-\!\epsilon_{n{\bf k-q}}\!-\!(z_j\!-\!i\delta)\!{\rm sgn}\!(\epsilon_{n{\bf k-q}}\!-\!\epsilon_f)}. \label{eq:SGM} \nonumber \\ 
\end{eqnarray}

The spectral function is calculated by
\begin{eqnarray}
A({\bf k},\omega)=\frac{1}{\pi} \sum_{\alpha} \Biggl| {\rm Im} \frac{1}{\omega-(\epsilon_{\alpha {\bf k}} + \Delta \Sigma_{\alpha {\bf k}} (\omega) + \Delta)} \Biggr|, \label{Akw}
\end{eqnarray}
where $\Delta \Sigma_{\alpha {\bf k}} (\omega)\!=\!\langle \psi_{\alpha {\bf k}}| \Sigma({\bf r,r'},\omega)\!-\!v_{{\rm XC}}({\bf r})\delta({\bf r}\!-\!{\bf r'}) |\psi_{\alpha {\bf k}} \rangle$, and $v_{{\rm XC}}$ is the exchange correlation potential. A shift $\Delta$ is introduced to keep the electron density to be the same as that of KS, $N_{elec}$, i.e., $\int_{-\infty}^{\epsilon_f} d\omega \int d{\bf k} A({\bf k},\omega)=N_{elec}$.

\section{results and discussions}\label{sec:result}
Our density-functional calculations are based on Tokyo Ab-initio Program Package (TAPP) (Ref.~\onlinecite{TAPP}) with plane-wave basis sets, where we employ norm-conserving pseudopotentials~\cite{PP-1,PP-2} and generalized gradient approximation (GGA) for the exchange-correlation potential.~\cite{PBE96} The experimental structure of (TMTSF)$_2$PF$_6$ obtained by a neutron measurement~\cite{structure-TMTSF} at 20 K is adopted. The cutoff energies in wavefunction and in charge densities are 36 Ry and 144 Ry, respectively, and an 11$\times$11$\times$3 $k$-point sampling is employed. The maximally localized Wannier function (MLWF) (Refs.~\onlinecite{MaxLoc-1} and \onlinecite{MaxLoc-2}) is used for interpolation of the self-energy and spectral function to a finer $k$ grid. The cutoff of polarization function is set to be 3 Ry and 198 bands are considered, which cover an energy range from the bottom of the occupied states near $-$30 eV to the top of the unoccupied states near 15 eV
 .
The integral over the Brillouin-zone (BZ) is evaluated by the generalized tetrahedron method.~\cite{Fujiwara,Nohara} The polarization up to $\omega$=86 eV is calculated in a logarithmic mesh with 110 energy points. The frequency dependence of the self-energy for the states near the Fermi level is calculated for [$-$30 eV: 30 eV] with the interval of 0.01 meV. The broadening $\delta$ in Eqs.~(\ref{eq:chi}), (\ref{eq:eps}) and (\ref{eq:SGM}) is set to 0.02 eV. The self energy at ${\bf q}$=${\bf G}$=${\bf 0}$ is treated in the manner in Ref.~\onlinecite{Spencer}. 

\begin{figure}[htbp]
\vspace{0cm}
\begin{center}
\includegraphics[width=0.5\textwidth]{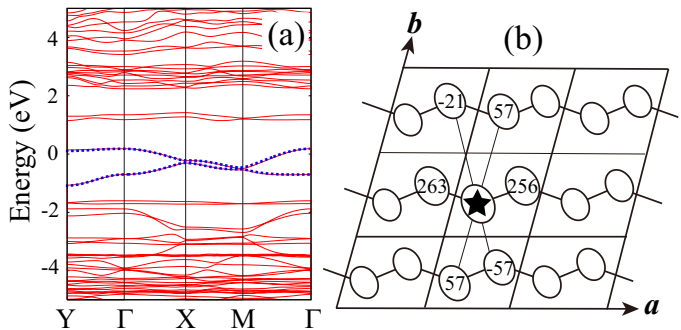}
\caption{(Color online) (a) GGA band structure of (TMTSF)$_2$PF$_6$, along the high-symmetry lines in the $ab$ plane, where $\Gamma$=(0, 0, 0), X=($a^{*}$/2, 0, 0), M=($a^{*}$/2, $b^{*}$/2, 0), Y=(0, $b^{*}$/2, 0). The Fermi level is at zero energy. (b) Schematic crystal structure in the $ab$ plane, where the unit cell contains two TMTSF molecules denoted by the ellipsoids. The number in the ellipsoids represent the transfer integral (in the unit of meV) between the highest-occupied molecular orbitals at the starred site and each site. The tight-binding bands [blue dotted curves in the panel (a)] are calculated from these transfers.} 
\label{Fig1}
\end{center}
\end{figure}

Figure~\ref{Fig1}~(a) shows the calculated GGA band structure of (TMTSF)$_2$PF$_6$. We find two narrow bands around the Fermi level, which are well separated in energy from other higher-energy bands. The appearance of such isolated low-energy bands is common to various organic conductors.~\cite{Nakamura-organic-1,Nakamura-organic-2,organic-1,organic-2,organic-3,organic-4,organic-5} We assign these bands to ``target bands" for which the self-energy effects are considered below. We construct MLWFs for the target bands and evaluate the transfer integrals as shown in the panel (b). The transfer integrals well reproduce the original bands as shown by the blue-dotted curves in the panel (a). Note that the transfers along the $a$ axis are about four times larger than those along the $b$ axis, reflecting a quasi-one-dimensional structure of the compound.

\begin{figure}[htbp]
\vspace{0cm}
\begin{center}
\includegraphics[width=0.4\textwidth]{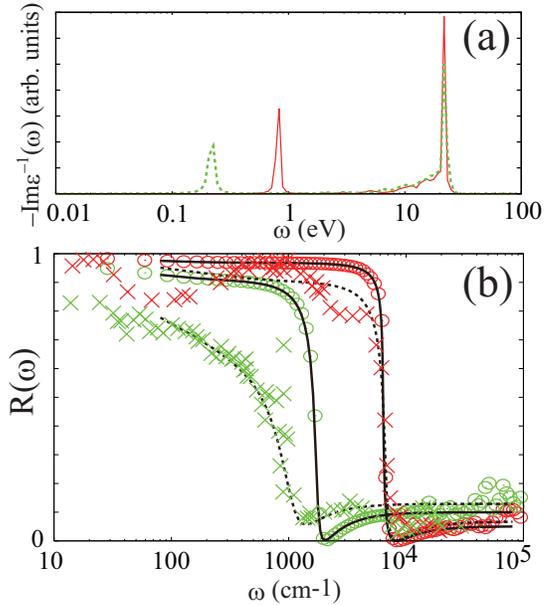}
\caption{(Color online) (a) Calculated energy loss function of (TMTSF)$_2$PF$_6$. The red-solid and green-dotted curves represent the spectra for $E\|a$ and $E\|b'$, respectively. (b) {\it Ab initio} reflectivity (circles) and experimental one (crosses) at 25 K (Ref.~\onlinecite{Jacobsen}). The results for $E\|a$ and $E\|b'$ are displayed by dark red and light green, respectively. The solid and dotted curves are the fits by Eq.~(\ref{Drude-model}) with parameters shown in Table \ref{PARAM-Drude}.}
\label{Fig2}
\end{center}
\end{figure}

Figure \ref{Fig2} (a) shows the calculated energy loss function $-{\rm Im}\epsilon^{-1}(\omega)$. The red-solid and green-dotted curves are results for the light polarization $E$ parallel to $a$ (along $x$ axis) and $b'$ (along $y$ axis), respectively. In the low-energy region, we find two plasmon peaks, at $\sim$0.2 eV for $E\|b'$ and $\sim$1.0 eV for $E\|a$. These peaks result from the plasmon excitations within the isolated target bands which have a low carrier density and small band widths. These plasmons are distinct from the plasmon seen at around 20 eV, which is relevant to the total charge density and the bare electron mass. We note that the difference between the plasmon peaks for $E\|a$ and $E\|b'$ reflects the difference in the transfer integrals ($\sim$260 meV along the $a$ axis and $\sim$60 meV along the $b$ axis). 

Figure \ref{Fig2} (b) compares the reflectance, 
\begin{eqnarray}
R_{\mu}(\omega)=\Biggl| \frac{1-\sqrt{\epsilon_{\mu}^{-1}(\omega)}}{1+\sqrt{\epsilon_{\mu}^{-1}(\omega)}} \Biggr|,  
\end{eqnarray}
between our theory (circles) and the experiment (crosses) of Ref.~\onlinecite{Jacobsen}. The calculated result reproduces the experimental one fairly well, in particular for $E\|a$ (dark red). The smaller energy scale of $E\|b'$ (light green) is also qualitatively reproduced. 

To quantify the comparison, we fit the following function to theoretical and experimental reflectance data  
\begin{eqnarray}
\epsilon_{\mu}(\omega)=\epsilon_{{\rm core,\mu}}-\frac{\omega_{pl,\mu\mu}^2}{\omega(\omega+i\delta)}.
\label{Drude-model}
\end{eqnarray}
For the theoretical data, $\omega_{pl,\mu\mu}$ is calculated with Eq.~(\ref{wpl}) and $\delta$ is fixed at 0.02 eV, so that the effective dielectric constant $\epsilon_{\rm core,\mu}$ is the only free parameter in the fitting. Table~\ref{PARAM-Drude} summarizes the values of $\epsilon_{{\rm core},\mu}$ and $\omega_{pl,\mu\mu}$. The calculated results well reproduce the experimental ones,~\cite{Note_delta} although the theoretical plasma edge of $E\|b'$ is by $\sim$1.5 times higher than that of experiment.  
\begin{table}[h] 
\caption{Comparison of Drude-model parameters in Eq.~(\ref{Drude-model}) for (TMTSF)$_2$PF$_6$ between theory and experiment of Ref.~\onlinecite{Jacobsen}. The unit of the plasma frequency is cm$^{-1}$.} 

\ 

\centering 
\begin{tabular}{lc@{\ \ \ }c@{\ \ \ }c@{\ \ \ }c@{\ \ \ }c} \hline \hline \\ [-8pt]  
 & \multicolumn{2}{c}{$E\|a$}  & 
 & \multicolumn{2}{c}{$E\|b'$} \\ \hline \\ [-8pt]  
 & $\epsilon_{{\rm core}}$ & $\omega_{pl}$ & 
 & $\epsilon_{{\rm core}}$ & $\omega_{pl}$  \\ \hline \\ [-8pt] 
 Theory & 2.5 & 10074 &  & 3.7 & 3331 \\  
 Expt.  & 2.9 & 11400 &  & 4.5 & 2360 \\ \hline \hline
\end{tabular} 
\label{PARAM-Drude} 
\end{table}

\begin{figure}[htbp]
\vspace{0cm}
\begin{center}
\includegraphics[width=0.5\textwidth]{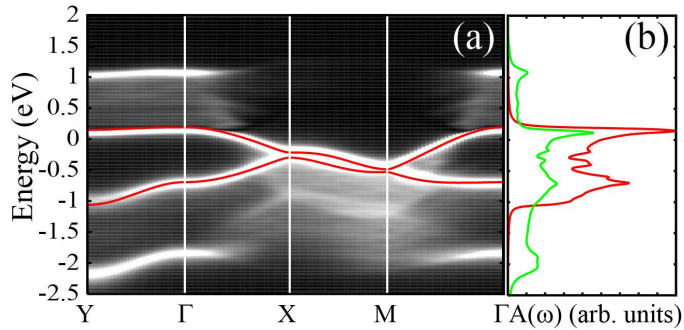}
\caption{(Color online) (a) Calculated spectral function $A({\bf k},\omega)$ of (TMTSF)$_2$PF$_6$, superposed by the Kohn-Sham band structure (red curves). The Fermi level is at zero energy. (b) Density of states obtained by KS (red-solid curve) and GW (green-dotted one).
}
\label{Fig3}
\end{center} 
\end{figure}

The low-energy plasmons found in Fig.~\ref{Fig2} can affect the low-energy electronic structure. To see this effect, we show in Fig.~\ref{Fig3} (a) the GW spectral function $A({\bf k},\omega)$ [Eq.~(\ref{Akw})]. While the quasiparticle band structure around $-$1.0-0.1 eV is similar to that of KS (red solid curves), we see an appreciable weight transfer to higher energy due to the self-energy effects. Along the Y-$\Gamma$ line, the new states emerge around by 1 eV above the unoccupied states, and around by 1 eV below the occupied states. Along the X-M line, the spectra are more broadened and spread in the range from $-$1.5 to 0 eV. In the panel (b), the density of states calculated by KS (red-solid curve) and by GW (green-dotted one) is displayed, from which we 
see that a considerable amount of weight is transferred to higher energy due to the self-energy effect.

\begin{figure}[htbp]
\vspace{0cm}
\begin{center}
\includegraphics[width=0.5\textwidth]{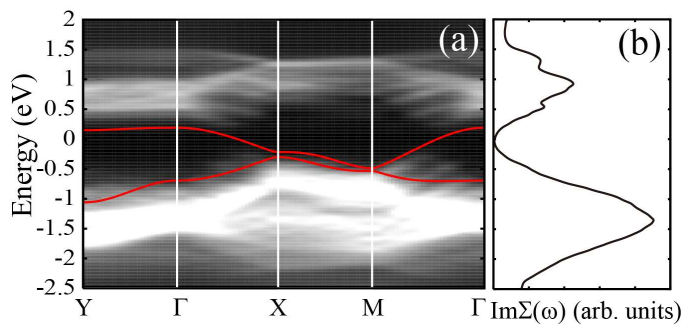}
\caption{(Color online) Calculated spectrum of (a) ${\rm Im}\Sigma({\bf k},\omega)$ and (b) ${\rm Im}\Sigma(\omega)$ of (TMTSF)$_2$PF$_6$. The Fermi level is at zero energy. The Kohn-Sham band structure (red solid curves) is superposed in the panel (a).}
\label{Fig4}
\end{center}
\end{figure}

To get insight into the relation between the dielectric function (Fig.~\ref{Fig2}) and the spectral function (Fig.~\ref{Fig3}), we plot in Fig.~\ref{Fig4} ${\rm Im}\Sigma({\bf k},\omega)$=$\sum_{\alpha}|{\rm Im}\Sigma_{\alpha{\bf k}}(\omega)|$ [(a)] and ${\rm Im}\Sigma(\omega)$=$\int d{\bf k}{\rm Im}\Sigma({\bf k},\omega)$ [(b)]. Im$\Sigma({\bf k},\omega)$ is directly related to the dielectric function through Eq.~(\ref{GW_SIGMA}), and to the spectral function through Eq.~(\ref{Akw}). We see that Im$\Sigma({\bf k},\omega)$ has strong intensities at 0.5$\sim$1.0 eV and $-$2.0$\sim$$-$0.5 eV; about 0.5 eV above the unoccupied states and about 0.5 eV below the occupied states. The energy scale of 0.5 eV roughly corresponds to the average of $\omega_{pl,b'}$$\sim$$0.2$ eV and $\omega_{pl,a}$$\sim$$1.0$ eV. Since the plasmons are known to make a peak in Im$\Sigma$ at energy $\epsilon_{unocc}$+$\omega_{pl}$ or $\epsilon_{occ}$$-$$\omega_{pl}$,~\cite{Aryasetiawan} the bright region 
 in Fig.~\ref{Fig4}(a) can be interpreted as an effect of the low-energy plasmons in Fig.~\ref{Fig2}, although the electron-electron scattering other than the plasmon excitation can also contribute to Im$\Sigma({\bf k},\omega)$. Since the strong peak of Im$\Sigma({\bf k},\omega)$ causes a large variation of Re$\Sigma({\bf k},\omega)$ through the Kramers-Kronig relation, new poles of the Green's function can be created just outside of the peak of Im$\Sigma({\bf k},\omega)$. This is consistent with $A({\bf k},\omega)$ in Fig. \ref{Fig3}.

\section{Conclusion}\label{sec:conclusion}

In summary, we study low-energy dynamical properties of an organic compound (TMTSF)$_2$PF$_6$ from first principles. Theoretical reflectance reproduces experimentally-observed plasma edges, and their anisotropy due to the quasi-one dimensional nature. The low-energy plasmons come out from the energetically-isolated bands around the Fermi level. The self-energy effect due to these plasmon excitations on the low-energy electronic structure is studied within the GW approximation. We have found that the self-energy effect is appreciable at energy by $\sim$0.5 eV above unoccupied states and below occupied states, suggesting that the plasmons can influence low-energy physics. Since organic conductors, or more generally, strongly-correlated electron materials, often have such isolated bands around the Fermi level, we expect that similar low-energy plasmon excitation can be relevant to physical properties of these materials. A detection of these effects in experiments such as photoem
 ission spectroscopy is an interesting future issue. 

\begin{acknowledgements} 
We would like to thank Takahiro Ito, Kyoko Ishizaka, Yoshiro Nohara, Yoshihide Yoshimoto, and Yusuke Nomura for useful discussions. Calculations were done at Supercomputer center at Institute for Solid State Physics, University of Tokyo. This work was supported by Grants-in-Aid for Scientific Research (No.~22740215, 22104010, 23110708, 23340095, 23510120, 25800200) from MEXT, Japan. 
\end{acknowledgements}


\begin{thebibliography}{999}
\bibitem{Note_delta}
The value of experimental $\delta$ at 25 K is 0.071 eV for $E\|a$ and 0.074 eV for $E\|b'$, which seems to be highly temperature dependent.~\cite{Jacobsen}
\end{thebibliography}
\end{document}